\documentclass[prb,twocolumn]{revtex4}
\usepackage{graphicx}
\usepackage{hyperref}
\usepackage{amssymb}
\usepackage{dcolumn}
\usepackage{float}
\usepackage{booktabs}
\usepackage{amsmath} % or simply amstext
\usepackage[utf8]{inputenc}
\usepackage{lmodern}
\usepackage{color}
\definecolor{red}{rgb}{1.0,0.0,0.0}
\usepackage{makecell}
\usepackage{marginnote}

\newcommand{\e}{\mathrm{e}}

\DeclareMathAlphabet{\bi}{OML}{cmm}{b}{it}

\def\ba{\begin{aligned}}
\def\ea{\end{aligned}}
\def\be{\begin{equation}}
\def\ee{\end{equation}}
\def\bearr{\begin{eqnarray}}
\def\eearr{\end{eqnarray}}

\usepackage{tikz}

\begin{document}
\title{Polarization and third-order Hall effect in III-V semiconductor heterojunctions}
%\title{Third-order Hall effect in two-dimensional charge carriers in semiconductor heterojunctions}
\bigskip
\author{Ojasvi Pal and Tarun Kanti Ghosh\\
\normalsize
Department of Physics, Indian Institute of Technology-Kanpur,
Kanpur-208016, India}
%\date{\today}
\begin{abstract}
We study Berry connection polarizability (BCP) induced electric polarization and third-order Hall (TOH) effect in a two-dimensional electron/hole gas (2DEG/2DHG) with Rashba-Dresselhaus (RD) spin-orbit couplings in III-V semiconductor heterostructures. The electric polarization decreases with increase of the Fermi energy and is responsive to the electric field orientation in the presence of RD spin-orbit couplings for both the systems.
We determine the BCP-induced TOH conductivity ($\chi_{\perp}^{\text{I}}$) along with the TOH conductivity associated with the band velocity ($\chi_{\perp}^{\text{II}}$). 
We find that the presence of an infinitesimal amount of Dresselhaus coupling in addition to the dominant Rashba coupling results in finite TOH responses. 
These conductivities vanish when the field is aligned with and/or orthogonal to the symmetry lines $k_x\pm k_y=0$ in both the systems.  
For typical system parameters in a 2DEG with $k$-linear RD interactions, the magnitude of $\chi_{\perp}^{\text{I}}$ is smaller than that of $\chi_{\perp}^{\text{II}}$. 
On the other hand, when both the SO couplings are comparable, 
$\chi_{\perp}^{\text{I}}$ shows a notable increase in magnitude, owing to the distinctive characteristics of BCP. 
The TOH conductivity of 2DEG remains unchanged when Rashba and Dresselhaus spin-orbit couplings 
are exchanged. For 2DHG with $k$-cubic RD interactions, $\chi_{\perp}^{\text{I},h}$ exhibits a larger magnitude compared to $\chi_{\perp}^{\text{II},h}$. Unlike the electron case, the BCP induced $\chi_{\perp}^{\text{I},h}$ alters under the exchange of spin-orbit 
coupling parameters, whereas $\chi_{\perp}^{\text{II},h}$ remains the same.

\end{abstract}

\maketitle
\section{Introduction}
The discovery of the Hall effect\cite{hall} in 1879 signified a crucial milestone in the field of condensed matter physics, paving the way for numerous notable advancements, such as the quantum Hall effect\cite{qhall}, the anomalous Hall effect\cite{nah,xiao}, the spin Hall effect\cite{she1,she2}, and the valley Hall effect\cite{valley}. The linear anomalous (conventional) Hall effect refers to the emergence of a transverse voltage in response to an applied electric current in the absence (presence) of  a magnetic field. In particular, the occurrence of the linear anomalous Hall effect relies on the broken time-reversal symmetry, which arises from intrinsic magnetic ordering within the system. These transport properties are substantially influenced by the Berry curvature, a geometrical property of the electronic wave function\cite{berry}. 

     Moreover, in the recent work of Sodemann and Fu, it has been proposed that time-reversal symmetric and noncentrosymmetric materials can exhibit second-order nonlinear Hall response which is mediated by the Berry curvature dipole moment\cite{fu}. It has been observed experimentally in layered transition metal dichalcogenides\cite{expt1,expt2,expt3}, which has subsequently propelled further investigations into other nonlinear related transport phenomena\cite{nl1,nl2,nl3,nl4,nl5,nl6}.
     
     In nonmagnetic materials characterized by inversion symmetry, the third-order Hall (TOH) response can prevail as the dominant effect, as both the linear anomalous Hall effect and second-order nonlinear Hall effect are absent in such systems. Gao \textit{et al}. introduced a semiclassical theory that incorporates second-order accuracy in external fields. Within this framework, they identified that the third-order Hall effect is induced by a geometric quantity known as the Berry connection polarizability (BCP)\cite{gao1,gao2}. The BCP is a second-rank tensor that quantifies the change in the field-induced Berry connection resulting from an applied electric field. Such extrinsic TOH response has recently been studied in a 2D Dirac model\cite{liu-prb}, the surface states of a hexagonal warped topological insulator\cite{tanay,awadhesh}.
% and Rashba systems with hexagonal warping\cite{awadhesh}. 
Experimental observations  have been reported in thick T$_d$-MoTe$_2$ samples\cite{liu}, few-layer WTe$_2$ flakes\cite{liao}, and the Weyl semimetal TaIrTe$_4$\cite{cwang}. Very recent studies also investigated the intrinsic 	TOH responses\cite{wang,kamal}.
     
Expanding upon recent research conducted on TOH within the realm of 2D Dirac materials with
tilted Dirac cone or trigonal warping term, here we study the TOH effect in electron and hole gases with  Rashba-Dresselhaus spin-orbit interaction (RSOI and DSOI) formed at the III-V semiconductor heterostructures. The RSOI emerges from the structure inversion asymmetry due to the confining potential, while the DSOI is a consequence of bulk inversion asymmetry. 
The transport properties such as electrical conductivity\cite{john,vasil,tkg,culcer-hole,culcer-hole1,kozlov,tkach}, spin Hall effect\cite{shen,sinova,chang,john1,zarea,mireles,firoz}, spin-galvanic photocurrent\cite{petra}, anomalous Hall effect\cite{bry,li}, magnetoplasmon\cite{plasmon}, optical conductivity\cite{alestin,carbotte} and zitterbewegung \cite{tutul,tutul2} has been studied extensively for the charge carriers at the semiconductor heterojunctions. The absence of Berry curvature in these systems prohibits both first and second-order Hall effects, emphasizing the significance of the TOH response. We find that an in-plane electric field induces electric polarization which is related to the BCP and the TOH response appears as the leading contribution in both the systems.

%Our theoretical study provide the detailed understanding of BCP which can be used to probe the TOH signals experimentally.

This paper is structured as follows: In Sec. \ref{section2}, we present the general formalism to calculate the electric polarization and TOH response within the framework of second-order semiclassical Boltzmann theory. In Sec. \ref{section3}, we initiate with a discussion on a 2DEG with $k$-linear RSOI and DSOI. Subsequently, we analyze the results of electric polarization and the transverse third-order conductivity. In Sec. \ref{section4}, we present the ground-state properties and BCP tensors of a 2DHG with $k$-cubic RSOI and DSOI and provide a discussion covering various aspects of the electric polarization and the transverse third-order conductivity of the system. Finally, we conclude and summarize our main results in Sec. \ref{section5}.

\section{Theoretical formulation}\label{section2}
In this section, we outline the general formalism to evaluate the electric polarization and  third-order Hall conductivity resulting from BCP in the absence of an external magnetic field. This formalism is based on the Boltzmann transport framework, employing the relaxation time approximation. The total current density is defined as
\begin{equation}\label{j}
{\mathbf{j}} = q \sum_{\lambda}\int[d\mathbf{k}]{\dot{\mathbf{r}}}^{\lambda}{f}^{\lambda}_{\mathbf{k}},
\end{equation} 
where $q$ is the charge of the current carriers, $[d\mathbf{k}]={d^2}k/{(2\pi)}^2$, ${f}^{\lambda}_{\mathbf{k}}$ denotes the non-equilibrium distribution function (NDF). The summation over $\lambda$ indicates the sum over different bands. Gao et al. developed a second-order semiclassical theory to calculate the third-order current response to an electric field\cite{gao1}. In this theory, the perturbation caused by an uniform electric field $\mathbf{E}$ is described as $H_E= - q \mathbf{E}\cdot{\mathbf{r}}$, resulting in a positional shift of the wavepacket. The semiclassical equations of motion incorporating the second-order corrections in electric field can be written as\cite{gao1,gao2}
\begin{equation}\label{eom}
\mathbf{\dot r}^{\lambda}=\frac{1}{\hbar}\left(\frac{\partial{\tilde{\epsilon}_{\lambda}}}{\partial{\mathbf{k}}}\right)-\mathbf{\dot k} \times  \tilde{\boldsymbol{\Omega}}_{\lambda}
\;\hspace{0.4cm} \text{and} \hspace{0.4cm}
\hbar \mathbf{\dot k} = q \mathbf{E}.
\end{equation}
To account for $\dot{\mathbf{r}}\propto E^2$, the band energy $\tilde{\epsilon}_{\lambda}$ is corrected to second order in $E$, while the Berry curvature $\tilde{\boldsymbol{\Omega}}_{\lambda}$ is corrected to first order in $E$. These corrections can be expressed as 
\begin{equation}
\tilde{\epsilon}_{\lambda}={\epsilon}_{\lambda}+{\epsilon}_{\lambda}^{(1)}+{\epsilon}_{\lambda}^{(2)} \; \hspace{0.4cm}\text{and} \hspace{0.4cm}   \tilde{\boldsymbol{\Omega}}_{\lambda}=\boldsymbol{\Omega}_{\lambda}+\boldsymbol{\Omega}^{(1)}_{\lambda},
\end{equation}
where ${\epsilon}_{\lambda}$ and $ \boldsymbol{\Omega}_{\lambda} $ are the unperturbed band energy and Berry curvature, respectively. The first-order correction to band energy can be obtained as ${\epsilon}_{\lambda}^{(1)}=\langle{u_{\lambda}}\vert{H_E}\vert{u_{\lambda}}\rangle
= -q \mathbf{E}\cdot \mathbf{A}_{\lambda\lambda}$, where $\mathbf{A}_{\lambda\lambda}=\langle{u_{\lambda}}\vert{i{\boldsymbol{\nabla}}_{\mathbf{k}}}\vert{u_{\lambda}}\rangle$ is the intraband Berry connection with  $|u_{\lambda} \rangle$ the cell-periodic unperturbed Bloch eigenstate. We omit the term ${\epsilon}_{\lambda}^{(1)}$ in our further calculations due to its gauge-dependent nature. Additionally, it can also be shown that ${\epsilon}_{\lambda}^{(1)}=0$ in the wave-packet picture\cite{gao1,tanay}. This is similar to the linear Stark effect, implying that the intrinsic dipole
moment of the system is zero as expected\cite{sakurai}.

The second-order energy correction is given by 
\begin{equation}
\begin{aligned}
{\epsilon}_{\lambda}^{(2)}&=\sum_{\lambda'\neq\lambda}\frac{{\vert\langle{u_{\lambda'}}\vert{H_E}\vert{u_{\lambda}}\rangle\vert}^{2}}{{\epsilon}_{\lambda}-{\epsilon}_{\lambda'}}\\
&\equiv q^2\sum_{\lambda'\neq \lambda}\frac{(\mathbf{E}\cdot{\mathbf{A}_{\lambda\lambda'}})(\mathbf{E}\cdot{\mathbf{A}_{\lambda'\lambda}})}{{\epsilon}_{\lambda}-{\epsilon}_{\lambda'}}.
\end{aligned}
\end{equation}
Here, ${\mathbf{A}_{\lambda\lambda'}}=\langle{u_{\lambda}}\vert{i{\boldsymbol{\nabla}}_{\mathbf{k}}}\vert{u_{\lambda'}}\rangle$ represents the interband Berry connection. The first-order correction to the Berry curvature is given by $\boldsymbol{\Omega}^{(1)}_{\lambda}={\boldsymbol{\nabla}}_{\mathbf{k}}\times\mathbf{A}_{\lambda}^{(1)}$, where $\mathbf{A}_{\lambda}^{(1)}$ corresponds to the first-order Berry connection. It can be expressed as  $\mathbf{A}_{\lambda}^{(1)}=\langle{u_{\lambda}^{(1)}}\vert i{\boldsymbol{\nabla}}_{\mathbf{k}}\vert{u_{\lambda}}\rangle + \text{c.c.}$, with the first-order correction to the eigenstate described as
\begin{equation}
\vert{u_{\lambda}^{(1)}}\rangle=\sum_{\lambda'\neq \lambda} 
\frac{-q(\mathbf{E}\cdot{\mathbf{A}_{\lambda'\lambda}})\vert{u_{\lambda'}}\rangle}{{\epsilon}_{\lambda}-{\epsilon}_{\lambda'}}.
\end{equation}
The field-induced Berry connection effectively captures the band geometric quantity, BCP and takes the form
\begin{equation}
{A}_{\lambda,a}^{(1)}=G_{ab}^{\lambda}E_b,
\end{equation}
where indices $a$ and $b$ denote the cartesian coordinates and the BCP tensor is defined as\cite{liu}
\begin{equation}\label{bcp}
G_{ab}^{\lambda} = - 2q  \; \text{Re}\sum_{\lambda'\neq \lambda}\frac{({{A}_{\lambda \lambda',a}})({{A}_{\lambda'\lambda,b}})}{{\epsilon}_{\lambda}-{\epsilon}_{\lambda'}}.
\end{equation}
Under an in-plane electric field, the second-order energy correction can be expressed in terms of BCP tensor as 
\begin{equation}\label{e2}
\epsilon_{\lambda}^{(2)}=-\frac{q}{2}(G_{xx}^{\lambda} E_{x}^{2}+2G_{xy}^{\lambda} E_{x}{E_y}+G_{yy}^{\lambda} E_{y}^{2}).
\end{equation}

It is to be noted that Eq. (\ref{e2}) resembles to the second-order Stark effect\cite{sakurai}.
So the BCP-induced dipole moment can be defined as
$D_{\lambda}({\bf k}) = -  \partial{\epsilon}_{\lambda}^{(2)}/\partial{E}$.
Quantum mechanically, the total electric polarization would be the sum over the polarizations 
of the occupied states in all the bands\cite{vanderbilt}.
Thus the electric polarization of a 2D system at zero-temperature can be expressed as 
\be
P=\sum_{\lambda} \int [d\mathbf{k}] D_{\lambda}({\bf k}).
\ee 
%where $D_{\lambda}({\bf k}) = - \frac{1}{2} \frac{ \partial{\epsilon}_{\lambda}^{(2)} }{\partial{E}}$ 
%is the induced dipole moment due to BCP. 
The electric polarization 
is simply the surface integral of BCP over all the occupied states in $k$-space. 
For an in-plane electric field ${\mathbf E} = E(\cos \theta, \sin \theta, 0)$,  
the electric polarization of a system can be written as 
\be\label{pol}
P=q E\sum_{\lambda}\int [d\mathbf{k}](G_{xx}^{\lambda}\cos^2\theta+G_{xy}^{\lambda}\sin2\theta+G_{yy}^{\lambda} \sin^2\theta).
\ee

%Having obtained the equations of motion accurate to the second-order in fields, we 
Next we move to calculate the NDF as a prerequisite for calculating the current. The Boltzmann transport equation within the relaxation time approximation to evaluate the NDF $f_{\mathbf{k}}^{\lambda}$ is given by\cite{ash}
\begin{equation}\label{BTE}
\dot{\mathbf{k}} \cdot\boldsymbol{\nabla}_{\mathbf{k}}f_{\mathbf{k}}^{\lambda} =-\frac{f_{\mathbf{k}}^{\lambda}-\tilde{f}_{\textnormal{eq}}^{\lambda}}{\tau}.
\end{equation}
The NDF can be obtained as 
\begin{equation}\label{NDF}
{f}_{\mathbf{k}}^{\lambda}=\sum_{\eta=0}^{\infty}\left(\frac{-q\tau}{\hbar}\mathbf{E}\cdot{\boldsymbol{\nabla}}_{\mathbf{k}}\right)^{\eta}\tilde{f}_{\text{eq}}^{\lambda}.
\end{equation}
Here, $\tau$ is the relaxation time and the Fermi-Dirac distribution function is given by
 $\tilde{f}_{\textnormal{eq}}^{\lambda} = 1/[1+ e^{\beta(\tilde{\epsilon}_{\lambda}-\mu)}]$. The distribution function encompasses $E$-dependence resulting from the band energy, accurate up to second-order in the electric field. One can expand it as  $\tilde{f}_{\text{eq}}^{\lambda}={f}_{\text{eq}}^{\lambda}+\epsilon_{\lambda}^{(2)}{f}_{\text{eq}}^{\prime\lambda}$, where ${f}_{\text{eq}}^{\lambda} $ is the equilibrium  distribution function defined in the absence of external electric field and ${f}_{\text{eq}}^{\prime \lambda}\equiv\partial{{f}_{\text{eq}}^{\lambda}}/{\partial{\epsilon}_{\lambda}}$. 

  We can derive the current by substituting the expressions of $\dot{\mathbf{r}}^{\lambda}$ and ${f}^{\lambda}_{\mathbf{k}}$ from Eqs. (\ref{eom}) and (\ref{NDF}), respectively into Eq. (\ref{j}). To obtain the third-order current, we collect the terms proportional to $E^3$, resulting in the following form\cite{liu-prb}
\begin{widetext}
\begin{equation}\label{current}
\begin{aligned}
\centering
\mathbf{j}^{(3)}&=-\frac{q^2}{\hbar}\sum_{\lambda}\int [d\mathbf{k}](\mathbf{E}\times\boldsymbol{\Omega}_{\lambda})[{\epsilon}_{\lambda}^{(2)}f_{\text{eq}}^{\prime \lambda}]-\frac{q^2\tau}{\hbar^2}\sum_{\lambda}\int [d\mathbf{k}]\Big\{\boldsymbol{\nabla}_{\mathbf{k}}\epsilon_{\lambda}(\mathbf{E}\cdot {\boldsymbol{\nabla}}_{\mathbf{k}})[{\epsilon}_{\lambda}^{(2)}f_{\text{eq}}^{\prime \lambda}]+\Big[-q(\mathbf{E}\times\boldsymbol{\Omega}^{(1)}_{\lambda})+\boldsymbol{\nabla}_{\mathbf{k}}\epsilon_{\lambda}^{(2)}\Big]\\
&(\mathbf{E}\cdot {\boldsymbol{\nabla}}_{\mathbf{k}})f_{\text{eq}}^{\lambda}\Big\}-\frac{q^4\tau^2}{\hbar^3}\sum_{\lambda}\int [d\mathbf{k}](\mathbf{E}\times\boldsymbol{\Omega}_{\lambda}){(\mathbf{E}\cdot {\boldsymbol{\nabla}}_{\mathbf{k}})}^{2}f_{\text{eq}}^{\lambda}
-\frac{q^4\tau^3}{\hbar^4}\sum_{\lambda}\int [d\mathbf{k}]\boldsymbol{\nabla}_{\mathbf{k}}\epsilon_{\lambda}{(\mathbf{E}\cdot {\boldsymbol{\nabla}}_{\mathbf{k}})}^{3}f_{\text{eq}}^{\lambda}.
\end{aligned}
\end{equation}
\end{widetext}
In the context of relaxation time, the first term, which is independent of $\tau$, and the term proportional to $\tau^2$ are both odd under time-reversal symmetry, causing them to vanish completely. Thus the third-order current exhibits dependencies on both $\tau$ and $\tau^3$. In Eq. (\ref{current}), the second term is attributed to the second-order energy correction within the distribution function, the third term arises due to the anomalous velocity generated by the first-order field correction to the Berry curvature and  the fourth term emerges as a consequence of the second-order field correction to the band velocity. Finally, the last term originates from the gradient term in the distribution function, which is cubic in the field. The third-order current response can be characterized as a Fermi surface property, as all the terms involved in its expression depend on the gradient of the equilibrium distribution function $f_{\text{eq}}$. We are interested in the third-order current induced by BCP, which is proportional to $\tau$, whereas $j^{(3)}\propto \tau^3$ is solely related to the band dispersion.

\begin{figure}[htbp]
\includegraphics[trim={0cm 0cm 0.0cm  0cm},clip,width=8.5
cm]{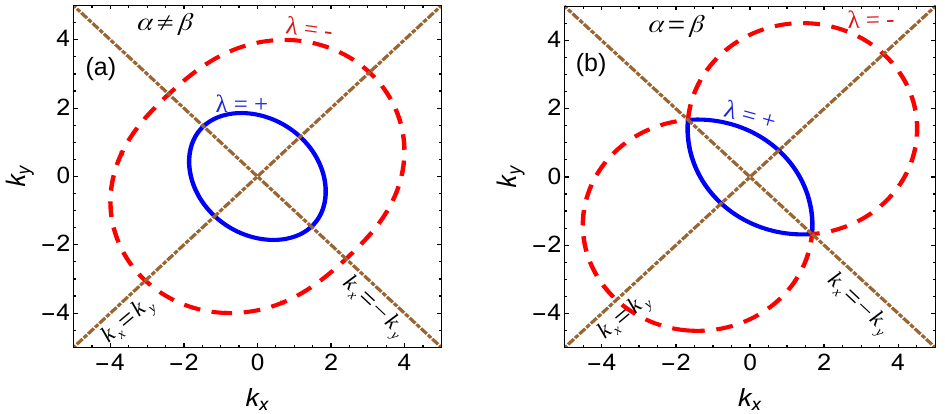}
\caption{Fermi contours along with the two mirror symmetric lines $k_y = \pm k_x$ of a 2DEG with $k$-linear Rashba-Dresselhaus spin-orbit couplings for
(a) $\alpha \neq \beta$ and (b) $\alpha = \beta$. Here, $k_x$ and $k_y$ are plotted in units of $k_0$. } 
\label{fig-contour-electron}
\end{figure}

\begin {figure*}
    \centering
    \includegraphics[width=1.0\textwidth]{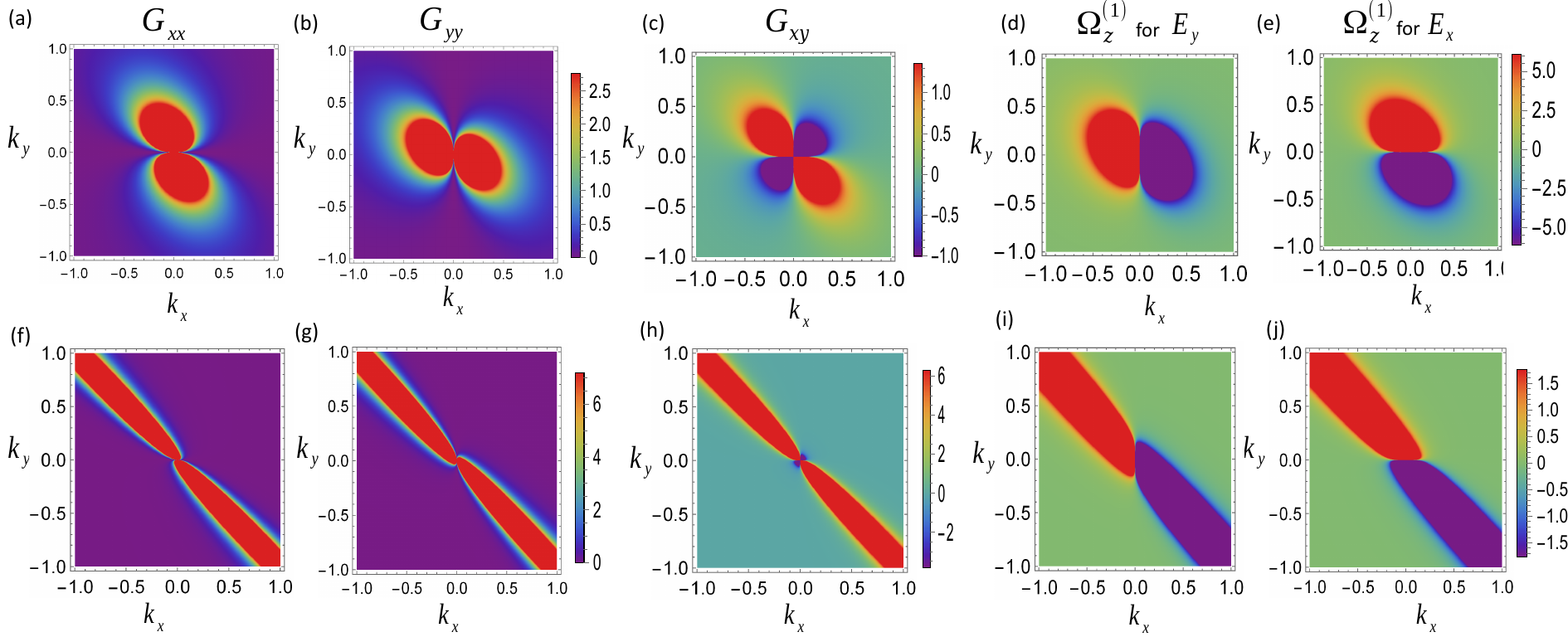}
\caption{We present density plots of the band geometric quantities and the field-induced Berry curvature of the 2DEG with linear Rashba-Dresselhaus spin-orbit interactions: (top panel) $\alpha= 6\times 10^{-9}$ eV cm and $\beta= 1\times 10^{-9}$ eV cm  and (bottom panel) $\alpha= 6\times 10^{-9}$ eV cm and $\beta= 5\times 10^{-9}$ eV cm.    
Here, (a)-(c) and (f)-(h) display the density plots of the BCP tensor components (in units of $e/\alpha {k_0^3}$),
%: (a) $G_{xx}$, (b) $G_{yy}$ and (c) $G_{xy}$. 
(d)-(e) and (i)-(j) display the field-induced Berry curvature 
$\Omega_z^{(1)}$ (in units of $ {e E/\alpha k_0^4}$) for two orientations of the electric field along the $y$ and  $x$ directions, respectively. 
The plots are given for the upper (+) band. 
We consider  $m_e= 0.024 m_0$, where $m_0$ is the free electron mass\cite{chang}. In both the panels, $k_x$ and $k_y$ are in units of $k_0$.} 
\label{fig1}
\end{figure*}

    The third-order current can be expressed in terms of third-order conductivity $\chi$ as
 $j^{(3)}_{a}=\chi_{abcd}E_b E_c E_d$, 
where the subscripts ${a,b ,c,d}\in\{x,y\}$ and $\chi_{abcd}$ is a rank-4 tensor. The third-order conductivity tensor comprises  two contributions, given by $\chi_{abcd} = \chi_{abcd}^{\text{I}} + \chi_{abcd}^{\text{II}}$, where $\chi_{abcd}^{\text{I}}$ is linear in $\tau$, and $\chi_{abcd}^{\text{II}}$ is proportional to $\tau^3$. These components can be derived as
\begin{equation}\label{tau}
\begin{aligned}
\chi_{abcd}^{\text{I}} & = \frac{-q^3\tau}{\hbar^2}\sum_{\lambda}\int [d\mathbf{k}] \Big\{\partial_{a}\partial_{b}G_{cd}^{\lambda}+\partial_{a}\partial_{d}G_{bc}^{\lambda}
-\partial_{b}\partial_{d}G_{ac}^{\lambda}\Big\}f_{\text{eq}}^{\lambda} \\ 
& + \frac{q^3\tau}{2}\sum_{\lambda}\int [d\mathbf{k}]v_{a,\lambda} v_{b,\lambda} G_{cd}^{\lambda}f_{\text{eq}}^{\prime\prime\lambda}
\end{aligned}
\end{equation}
and 
\begin{equation}\label{tau3}
\begin{aligned}
\chi_{abcd}^{\text{II}}&=-\frac{q^4\tau^3}{\hbar^3}\sum_{\lambda}\int [d\mathbf{k}] v_{a,\lambda}\partial_{b}\partial_{c}\partial_{d}f_{\text{eq}}^{\lambda}.
\end{aligned}
\end{equation}
Here, $\hbar\mathbf{v}_{\mathbf{k},\lambda}={\boldsymbol{\nabla}}_{\mathbf{k}}\epsilon_{\lambda}$ is the unperturbed band velocity. It is evident from Eqs. (\ref{tau}) and (\ref{tau3}) that  $\chi_{abcd}^{\text{I}}$ is associated with BCP, and $\chi_{abcd}^{\text{II}}$ is connected to the band dispersion. Next, we consider an in-plane electric field $\mathbf{E}=(E\cos\theta, E\sin\theta,0)$ such that the applied electric field forms an angle $\theta$ with respect to the $x$ axis. The third-order current within the plane can be described as 
\begin{equation}
\left (\begin{matrix}
  { {j}^{(3)}_x}\\
  \\
 {{j}^{(3)}_y}
\end{matrix} \right)= 
 \left (\begin{matrix}
  {{\chi_{11}}{E_x^3}+ 3\chi_{12}{E_x} {E_y^2}+3\chi_{13}{E_x^2}{E_y}+\chi_{14}{E_y^3}}\\
  \\
 {\chi_{41}{E_x^3}+3\chi_{31}{E_x}{E_y^2}+3\chi_{21}{E_x^2}{E_y}+{\chi_{22}}{E_y^3}}
\end{matrix} \right),
\end{equation}
where we define $\chi_{11}=\chi_{xxxx}$, $\chi_{12}=(\chi_{xyyx}+\chi_{xyxy}+\chi_{xxyy})/3$, $\chi_{13}=(\chi_{xxxy}+\chi_{xyxx}+\chi_{xxyx})/3$, $\chi_{14}=\chi_{xyyy}$, $\chi_{41}=\chi_{yxxx}$, $\chi_{31}=(\chi_{yyyx}+\chi_{yxyy}+\chi_{yyxy})/3$, $\chi_{21}=(\chi_{yxxy}+\chi_{yxyx}+\chi_{yyxx})/3$ and $\chi_{22}=\chi_{yyyy}$. In nonlinear transport experiments, one measures the current response that is transverse to the electric field. Therefore, our focus lies in the third-order transverse current, which can be written as $j^{(3)}_{\perp}(\theta)=\mathbf{j}^{(3)}\cdot(\hat{\mathbf{z}}\times\hat{\mathbf{E}})$ and the associated third-order transverse conductivity is defined as
%\begin{equation}
$ \chi_{\perp}(\theta)= j^{(3)}_{\perp}/E^3$.
%\end{equation}
For a 2D system, the explicit form of $\chi_{\perp}(\theta)$ is given by
\begin{widetext}
\begin{equation}\label{chi}
\chi_{\perp}(\theta)=(3\chi_{21}-\chi_{11})\cos^{3}\theta \sin\theta+(\chi_{22}-3\chi_{12}
)\sin^{3}\theta \cos\theta+3(\chi_{31}-\chi_{13})\cos^{2}\theta \sin^2\theta+\chi_{41}\cos^{4}\theta-\chi_{14}\sin^{4}\theta.
\end{equation} 
\end{widetext}
Importantly, the transverse third-order conductivities $\chi_{\perp}^{\text{I}}$ and $\chi_{\perp}^{\text{II}}$, proportional to $\tau$ and $\tau^3$, adopts the same form as that of $\chi_{\perp}$. This modification involves substituting 
$\chi_{abcd}$ by $\chi_{abcd}^{\text{I}}$ for $\chi_{\perp}^{\text{I}}$, and with $\chi_{abcd}^{\text{II}}$ for $\chi_{\perp}^{\text{II}}$. In the next section, we will apply this formalism to the Rashba-Dresselhaus system and investigate its third-order transverse conductivity.

\section{Two-dimensional electron gas with $k$-linear  Rashba-Dresselhaus spin-orbit coupling}\label{section3}
The Hamiltonian for a 2DEG with $k$-linear RSOI and DSOI is given by\cite{john,shen}
\begin{equation}
H=\frac{\hbar^2 k^2}{2 m_e}+\alpha(\sigma_x k_y-\sigma_y k_x)+\beta(\sigma_x k_x-\sigma_y k_y).
\end{equation}
Here, $\alpha$ and $\beta$ represent the strengths of RSOI and DSOI, $m_e$ denotes the effective mass of an electron and the $\sigma$'s are the Pauli matrices. 
%There exists a conserved quantity,  $(\sigma_x-\sigma_y)$, when $\alpha=\beta$ due to the commutation relation $[H,(\sigma_x-\sigma_y)]=0$\cite{john}. 
The energy spectrum consists of two bands ($\lambda=\pm$) of the following form
\begin{equation}
\epsilon_{\lambda}(\mathbf{k})=\frac{\hbar^2k^2}{2m_e}+\lambda\sqrt{{(\alpha k_y+\beta k_x)}^{2}+{(\alpha k_x+\beta k_y)}^{2}}.
\end{equation}
The corresponding eigenspinors can be obtained as $\vert{u}_{\lambda}\rangle=(1/\sqrt{2})[1 \hspace{0.25 cm}\lambda i \e^{i\varphi}]^{T} $, where $\varphi=\tan^{-1}[({\alpha k_y+\beta k_x})/({\alpha k_x+\beta k_y})]$ with $k_x=k \cos\phi$ and $k_y=k \sin\phi$ and $T$ being the transpose operation.

The two bands $ \epsilon_{\lambda}(\mathbf{k})$ meet at ${\mathbf k}=0$, commonly called a band touching point (BTP). The energy difference between the two bands is given by 
$ \epsilon_g(\mathbf{k}) = 2\Lambda_{\mathbf{k}}$  
with $\Lambda_{\mathbf{k}}\equiv\Lambda=\sqrt{{(\alpha k_y+\beta k_x)}^{2}+{(\alpha k_x+\beta k_y)}^{2}}$. 
The maximum value of $ \epsilon_g(\mathbf{k}) $ at  
$\phi = \pi/4$ and $5 \pi/4$ is $ 2k(\alpha + \beta)$, while 
the minimum value of $ \epsilon_g(\mathbf{k}) $ at $\phi =3\pi/4 $ and $7\pi/4$ is  
$ 2k|\alpha - \beta| $. These values of $\phi$ also coincide with
the symmetry lines $k_x \pm k_y = 0$ of the system. 
There is a line degeneracy along the symmetry line $k_y + k_x = 0$ for $\alpha = \beta $ case
 as shown Fig. \ref{fig-contour-electron}.

%The eigenspinors can be obtained as $\vert{u}_{\lambda}\rangle=(1/\sqrt{2})[1 \hspace{0.25 cm}\lambda i \e^{i\varphi}]^{T} $, where $\varphi=\tan^{-1}[({\alpha k_y+\beta k_x})/({\alpha k_x+\beta k_y})]$ with $k_x=k \cos\phi$ and $k_y=k \sin\phi$ and $T$ being the transpose operation. 
%The band velocities of the system can be obtained as
%\begin{equation}
%\begin{aligned}
%v_{x,{\lambda}}&=\frac{\alpha}{\hbar}\Bigg[\tilde{k}\cos\phi+\lambda\frac{(1+\zeta^{2})\cos\phi+2\zeta\sin\phi}{\sqrt{\gamma}}%\Bigg],\\
%v_{y,{\lambda}}&=\frac{\alpha}{\hbar}\Bigg[\tilde{k}\sin\phi+\lambda\frac{(1+\zeta^2)\sin\phi+2\zeta\cos\phi}{\sqrt{\gamma}}%\Bigg],
%\end{aligned}
%\end{equation}

       The wave vectors corresponding to $\epsilon >0$ are given by  $\tilde{k}_{\lambda}(\phi)=-\lambda\sqrt{\gamma}+\sqrt{2\tilde{\epsilon}+\gamma}$, where we define $\gamma=1+\zeta^2+2\zeta\sin2\phi$, with $\zeta=\beta/\alpha$.  We introduce the scaled parameters  $\tilde{k}=k/k_0$ and  $\tilde{\epsilon}=\epsilon/\epsilon_0$ with $k_0=m\alpha/\hbar^2$ and $\epsilon_0=m\alpha^2/\hbar^2$ as scaled wave vector and energy, respectively.  For $\epsilon <0$, only one energy band with $\lambda=-$ contributes and it attains a minimum value of $\tilde{\epsilon}_{\text{min}}=-\gamma/2$. The associated wave vectors can be expressed as $\tilde{k}_{\eta}(\phi)=\sqrt{\gamma}-{(-1)}^{\eta-1}\sqrt{2\tilde{\epsilon}+\gamma}$, where $\eta=1,2 $ is the branch index. 
%It is crucial to note that the presence of both couplings results in the angular anisotropy of the energy spectrum and Fermi contours. This anisotropy stems from the $\sin2\phi$ dependence in the $\gamma$ term. When either of the SOI couplings goes to zero, i.e., $\alpha=0$ or $\beta=0$, the energy spectrum becomes isotropic and the Fermi contours transform into concentric circles\cite{john}.

We consider the following key points in order to study the TOH response of this system. 
%The Berry phase acquired by a state can be obtained by evaluating the line integral  along a %closed loop that encloses the degenerate point, $\Gamma_{\lambda}=\oint d\boldsymbol{\ell} \cdot %\mathbf{A}_{\lambda \lambda}= -\pi{\text{sgn}(\alpha^2-\beta^2)}$.
%    Here, the values of $\Gamma_{\lambda}$ are $-\pi$ ($\pi$) when $\alpha^2$ is greater (smaller) than $\beta^2$ and they are zero when $\alpha=\beta$. 
The conventional Berry curvature of the system vanishes everywhere except for a singular nature at the degenerate point $\mathbf{k}=0$. As a result, the linear anomalous Hall effect and Berry curvature dipole induced second-order Hall response vanish. Hence, the BCP induced third-order Hall response will be the dominant one in the $k$-linear Rashba-Dresselhaus system.
To determine the third-order conductivity, one can compute the different components of the BCP tensor using Eq. (\ref{bcp}) as
\begin{equation}
G_{ab}^{\lambda}=\lambda\frac{ e {(\alpha^2-\beta^2)}^2}{4\Lambda^5}\left (\begin{matrix}
   {k_{y}^{2}}  & {-k_{x} k_{y}} \\
  
  {-k_{x} k_{y}} & {k_{x}^{2}}
\end{matrix} \right),
\end{equation}
%It is important to note that the $G_{ab}$ component becomes zero for $\alpha=\beta$. This implies that the third-order conductivity resulting from BCP can only be observed in the system when $\alpha\neq\beta$. 
We have plotted the density plots of  these BCP tensor elements $G_{xx}$, $G_{yy}$ and $G_{xy}$  for $\beta\ll\alpha$  in Figs. \ref{fig1}(a)-\ref{fig1}(c), respectively. The diagonal elements $G _{xx}$ and $G_{yy}$ exhibit a dumbbell-like pattern, whereas the off-diagonal element $G_{xy}$ shows a quadrupole-like structure. Under an in-plane electric field, the field-induced Berry curvature can be written in terms of BCP tensor as\cite{liu} $\boldsymbol{\Omega}^{(1)}_{\lambda}=[({\partial_{k_x}}G_{yx}^{\lambda}-{\partial_{k_y}}G_{xx}^{\lambda})E_x+({\partial_{k_x}}G_{yy}^{\lambda}-{\partial_{k_y}}G_{xy}^{\lambda})E_y ]\hat{\mathbf{z}}.$ We find that for this system, the second-order energy correction and field-induced Berry curvature can be obtained as
\begin{equation}\label{energy2}
\epsilon_{\lambda}^{(2)}=\lambda\frac{e^2{(\alpha^2-\beta^2)}^2}{8\Lambda^5}(\mathbf{E}\times\mathbf{k})^{2}
\end{equation}
and
\begin{equation}\label{bc1}
\boldsymbol{\Omega}^{(1)}_{\lambda}(\mathbf{k})=\lambda\frac{e{(\alpha^2-\beta^2)}^2}{2\Lambda^5}(\mathbf{E}\times\mathbf{k}).
\end{equation}
It should be mentioned here that expressions of $\epsilon_{\lambda}^{(2)}$ and
$\boldsymbol{\Omega}^{(1)}_{\lambda}(\mathbf{k})$ are obtained using the non-degenerate perturbation theory. Therefore Eqs. (\ref{energy2}) and (\ref{bc1}) are not valid at $\alpha=\beta $ case,
since there is a line degeneracy along the symmetry line $k_y + k_x= 0$ for $\alpha = \beta $ case.

Unlike the Berry curvature, the field-induced Berry curvature remains finite and exhibits a dipole-like structure. It is directed out-of-plane, but its orientation is sensitive to the applied electric field. Figures \ref{fig1}(f)-\ref{fig1}(j) further depict that as the values of $\alpha$ approach close to $\beta$, the lobes in the diagonal element of BCP and $\Omega_z^{(1)}$ undergo substantial elongation. In the case of $G_{xy}$, the lobes experience stretching in one direction, accompanied by a corresponding contraction in the orthogonal direction. Note that these BCP tensor elements and ${\Omega}_z^{(1)}$ are concentrated around the BTP. 
%degenerate point, i.e., small gap region. 
Figure \ref{fig1} clearly demonstrates that the lobes in diagonal components of BCP and ${\Omega}_z^{(1)}$  are confined in the $x$-$y$ plane. This observation can be understood from the system's anisotropic nature resulting from $\alpha\neq\beta$. For the pure Rashba system ($\beta=0$), the lobes are exclusively aligned along the $x$ and $y$ directions.
 \begin{figure}[htbp]
\includegraphics[trim={0cm 0cm 0.0cm  0cm},clip,width=8.5
cm]{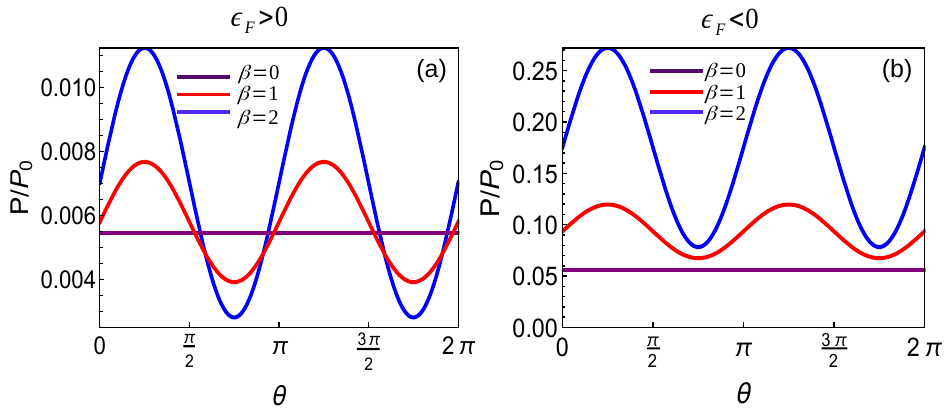}
\caption{Polarization (in units of $P_0=e^2 E/\epsilon_0$) as a function of angle $\theta$ 
for different  Dresselhaus coupling strengths (in units of $10^{-9}$ eV cm) at a fixed Rashba coupling strength of $\alpha = 6 \times 10^{-9}$ eV cm: (a) $\epsilon_F > 0$ at a fixed electron density of $n_e = 5.7 \times 10^{10}$/cm$^{2}$ and (b) $\epsilon_F < 0$ at $n_e = 10^{10}$/cm$^{2}$. The other parameters used are the same as in Fig. \ref{fig1}.  }  
\label{pol-electron}
\end{figure}
\begin {figure*}
    \centering
    \includegraphics[width=1\textwidth]{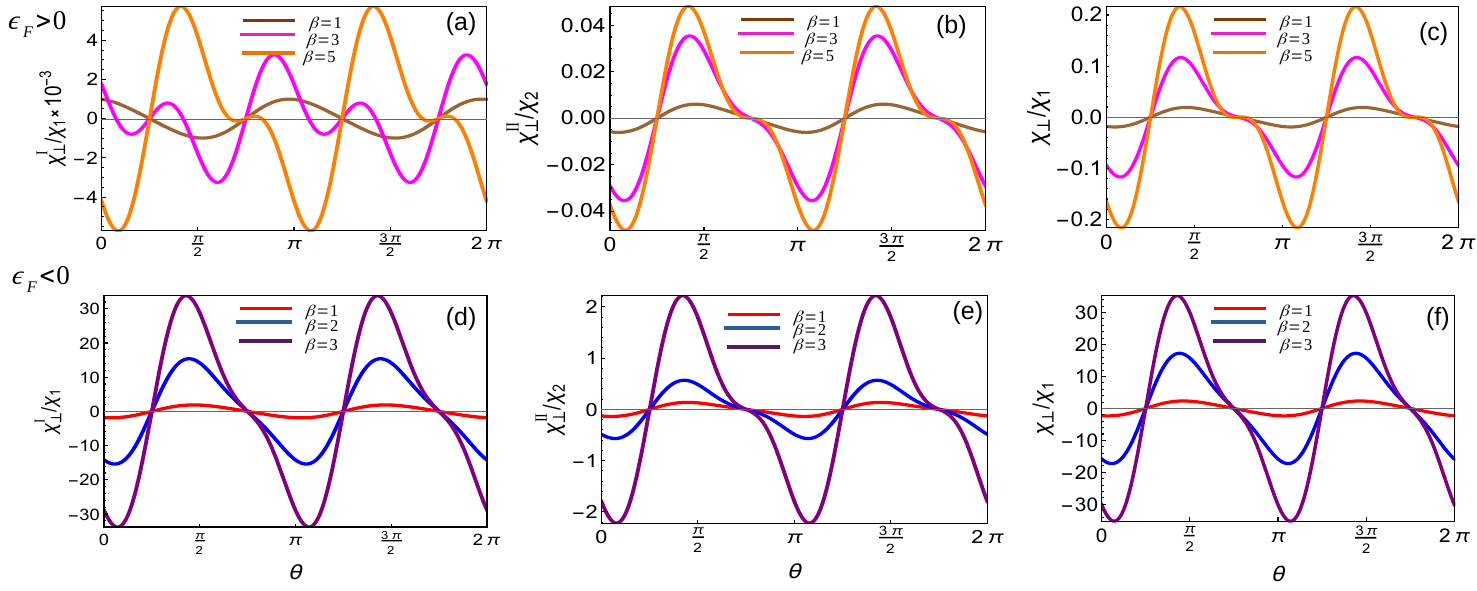}
 \caption{Variation of the transverse third-order conductivities with angle $\theta$ for different Dresselhaus coupling strengths, while keeping the Rashba coupling strength fixed at $\alpha = 6 \times 10^{-9}$ eV cm. The conductivities $\chi_{\perp}^{\text{I}}$ and $\chi_{\perp}^{\text{II}}$ correspond to transverse third-order conductivities proportional to $\tau$ and $\tau^3$, respectively. The total transverse conductivity is given by $\chi_{\perp} = \chi_{\perp}^{\text{I}} + \chi_{\perp}^{\text{II}}$. The top panel (a)-(c) represents the case for $\epsilon_F > 0$, and the bottom panel (d)-(f) corresponds to the scenario where $\epsilon_F < 0$. The normalization parameters for conductivities, $\chi_{\perp}^{\text{I}}$ and $\chi_{\perp}^{\text{II}}$, are given by $\chi_{1} = {\tau e^4 \hbar^4}/{m_e^3 \alpha^4}$ and $\chi_{2}={e^4 \tau^3}/{m_e \hbar^2}$, respectively. The value of  $\beta$ is given in units of $10^{-9}$ eV cm. The $\beta=5$ curve is scaled by factors of 50 in (a) and 5 in (b) and (c), while the $\beta=3$ curve is scaled down by factors of 5 in (d) and (f). The parameters used are same as in Fig. \ref{pol-electron}. }  
\label{fig2} 
\end{figure*}

\subsection{ Polarization}
%For an in-plane electric field, using Eq. (\ref{energy2}),  the polarization of a system can be written as $P=-eE\sum_{\lambda}\int [d\mathbf{k}](G_{xx}^{\lambda}\cos^2\theta+G_{xy}^{\lambda}\sin2\theta+G_{yy}^{\lambda} \sin^2\theta)$. 
For the pure Rashba system ($\beta=0$), an analytical expression of the electric polarization can be obtained using Eq. (\ref{pol}) as\begin{equation}\label{pol-e}
P=\frac{P_0}{16\pi } \begin{cases} \frac{1}{N_e-1}, & N_e >1 , \\ 
\frac{2N_e}{1-N_e^2}, & N_e < 1, \end{cases}
\end{equation}
where $P_0=e^2 E/\epsilon_0$ and 
$N_e = \pi l_e^2n_e $ with  
$ l_e = \hbar^2/(m_e \alpha) $.
Note that the Fermi energy is zero at the BTP which can be reached if $ N_e= \pi l_e^2 n_e =1$. 
%\begin{equation}\label{pol-e}
%P=\frac{P_0}{16\pi } \frac{1}{\tilde{n}_e}\begin{cases} 1, & {\epsilon}_F >0 , \\ -\sqrt{2{\tilde{n}}_e+1},& {\epsilon}_F <0, \end{cases}
%\end{equation}
%where $P_0=e^2 E/\epsilon_0$ and $\tilde n_e = \pi n_e l_e^2 $ with 
%$ l_e = \hbar^2/(m\alpha) $. 
For $\alpha=0$ but $\beta \neq 0$, the polarization can be obtained from Eq. (\ref{pol-e}) with
$\alpha $ replaced by $\beta$.
We find that the polarization decreases with the increase in Fermi energy, reflecting the behavior of the BCP. It is important to note that the polarization does not vary with the angle $\theta$ (between the electric field and $x$ axis) since contribution from $G_{xy}\propto k_xk_y$ vanishes upon angular integration, as $\int_{0}^{2\pi}\sin2\phi d\phi=0$. Both $G_{xx}$ and $G_{yy}$ contributes equally, rendering it insensitive to orientation of the electric field in the case of $\beta=0$.

   We have also illustrated the dependence of polarization on $\theta$ under the influence of both the couplings in Figs. \ref{pol-electron}(a) and \ref{pol-electron}(b) for ${\epsilon}_F > 0$ and ${\epsilon}_F < 0$, respectively. This demonstrates that adding an infinitesimal DSOI to the RSOI makes  polarization responsive to the electric field orientation, as  $G_{xy}$ also contributes. Therefore, the polarization takes the following form: $P=-eE\sum_{\lambda}\int [d\mathbf{k}](G_{xx}^{\lambda}+G_{xy}^{\lambda}\sin2\theta)$. The integration of $G_{xx}$ and $G_{xy}$ yields the positive values for the given set of parameters. Consequently, the polarization is maximum at $\theta=\pi/4$ and $5\pi/4$ and minimum at $\theta=3\pi/4$ and $7\pi/4$. These values of $\theta$ coincides with the symmetry lines of the system. The magnitude of  polarization increases with an increase in $\beta$ for a given $\alpha$. The electric polarization in ${\epsilon}_F < 0$ region is large as compared to
${\epsilon}_F > 0$. This is due to the Van Hove singularity in the density of states as 
Fermi energy approaches the band minimum, $\epsilon_F \rightarrow \epsilon_{\text{min}}$.
  
    %Now that we have evaluated the BCP tensor and the field-induced Berry curvature, we can proceed to determine the  third-order transverse conductivity of the system.  
   
\subsection{ Third-order transverse conductivity}
   In the Rashba-Dresselhaus system, where both $\alpha$ and $\beta$ are nonzero, the lines $k_x=\pm k_y$ serve as symmetry axes of the system. Due to the underlying symmetry axes of the system, we have $\chi_{11}=\chi_{22}$, $\chi_{12}=\chi_{21}$, $\chi_{31}=\chi_{13}$, and $\chi_{14}=\chi_{41}$, which reduces Eq. (\ref{chi}) to
\begin{equation}\label{chi-perp}
\chi_{\perp}(\theta)=\frac{1}{4}(3\chi_{21}-\chi_{11})\sin4\theta+\chi_{41}\cos{2}\theta.
\end{equation}
The vanishing behavior of  $\chi_{\perp}$ along or perpendicular to the  symmetry lines of the system can be understood well from the above equation. Both the terms $\sin 4 \theta $ and $\cos 2\theta$ of Eq. (\ref{chi-perp}) vanish simulatneously whenever $ \theta \in\left\{{\pi}/{4},{3\pi}/{4}, {5\pi}/{4}, {7\pi}/{4}\right\}$, independent of the system parameters. These four angles coincide with the symmetry lines of the systems. If we consider $\theta=0$, then $\chi_{\perp}=\chi_{41}=\chi_{yxxx}$. Below, we will discuss the contributions to transverse conductivity based on their scaling relation with $\tau$.

   \textit{\textbf{$\tau$- scaling conductivity}} ($\chi_{\perp}^{\text{I}}$): We numerically evaluate $\chi_{\perp}^{\text{I}}$ for the system, considering both $\epsilon_F > 0$ and $\epsilon_F < 0$. For the isotropic Rashba system ($\beta=0$), we observe that $\chi_{11}=3\chi_{21}$ and $\chi_{41}=0$. Consequently, $\chi_{\perp}^{\text{I}}$ vanishes for all Fermi energies. For Fermi energies above the BTP, we perform the calculations at a constant electron density of $n_e=5.7\times 10^{10}$/cm$^{2}$ and a fixed Rashba coupling strength of $\alpha = 6\times 10^{-9}$ eV cm, while systematically varying the Dresselhaus coupling parameter $\beta$. The variation of $\chi_{\perp}^{\text{I}}$ as a function of the angle $\theta$ for different values of $\beta$ is shown in Fig. \ref{fig2}(a). We find that when the value of $\beta$ is much smaller than $\alpha$, let's say $\beta=1$, we obtain a finite $\chi_{\perp}^{\text{I}}$ that exhibits significant dependence on the $\cos 4\theta$ term. The system exhibits more anisotropic behavior as we further increase $\beta$, a competition arises between the coefficients of $\sin4\theta$ and $\cos2\theta$, which is clearly illustrated in Fig. \ref{fig2}(a). We also observe the presence of additional angles $\theta$ at which $\chi_{\perp}^{\text{I}}$ vanishes.  Note that these angles of additional zeros depend on the system parameters. They  manifest symmetrically around the zeros that originate from the inherent symmetry of the system, i.e., $\theta=\pi/4, 3\pi/4, 5\pi/4, 7\pi/4$. Additionally, it can be noted that the magnitude of $\chi_{\perp}^{\text{I}}$ increases significantly as $\beta$ approaches close to $\alpha$ (as shown here for $\beta=5$). At $\alpha=\beta$, $\chi_{\perp}^{\text{I}}=0$. This behavior can be attributed to the characteristics of the BCP tensor. The variation of $\chi_{\perp}^{\text{I}}$ with $\theta$ exhibits a periodicity of $\pi$. On the other hand, the magnitude of $\chi_{\perp}^{\text{I}}$ for $\epsilon_F<0$ (${\epsilon}_{\text{min}}<\epsilon_F<0$) is notably larger compared to $\epsilon_F>0$, as depicted in Fig. \ref{fig2}(d). At Fermi energies below the BTP, the conductivity increases significantly as the Fermi energy approaches the band minimum, attributed to the Van Hove singularity in the density of states as $\epsilon_{F}\rightarrow\epsilon_{\text{min}}$.

  One can determine the maxima and minima of $\chi_{\perp}$ by differentiating Eq. (\ref{chi-perp}) with respect to $\theta$ and set it zero. Then we obtain locations of maxima and mimima for various system
parameters. The values of the coefficients of $\sin 4 \theta $ and $\cos 2\theta$ of Eq. (\ref{chi-perp}) change with $\alpha$ and $\beta$, leading to shifts in the positions of maxima and minima and emphasizing their dependence on system parameters.

   It is important to emphasize that the magnitude and sign of $\chi_{\perp}^{\text{I}}$ remain unaltered when the values of $\alpha$ and $\beta$ are interchanged. For instance, $\chi_{\perp}^{\text{I}}(\alpha=2, \beta=6$) = $\chi_{\perp}^{\text{I}}(\alpha=6, \beta=2$). This finding can be explained by the invariance of the Hamiltonian under $\alpha\leftrightarrow\beta$ and rotation by the unitary rotation operator,  $U=e^{-i\frac{\pi}{4}\sigma_z}e^{-i\frac{\pi}{2}\sigma_y}$, which transforms $\sigma_x\rightarrow-\sigma_y$, $\sigma_y\rightarrow-\sigma_x$, and $\sigma_z\rightarrow-\sigma_z$. Both the unperturbed velocity operator and the velocity resulting from the second-order energy correction, which is related to the BCP tensor, also remain invariant under these transformations. Thus the third-order current is same when $\alpha$ and $\beta$ are exchanged. 
   
    We also explore the dependence of third-order conductivity on the Fermi energy. Keeping the electron density and Rashba coupling $\alpha$ fixed, an increase in $\beta$ leads to a reduction in the Fermi energy. Consequently, we find that the magnitude of $\chi_{\perp}^{\text{I}}$ increases as the Fermi energy decreases. This understanding can be derived from the behavior of the BCP tensors, which exhibit a maximal value at the degenerate point and gradually decrease as one moves away from it.
    
  \textit{\textbf{$\tau^3$-scaling conductivity}} ($\chi_{\perp}^{\text{II}}$): We also evaluate the transverse third-order conductivity $\chi_{\perp}^{\text{II}}$, which is proportional to $\tau^3$ and solely arises from the band velocity. We find that $\chi_{\perp}^{\text{II}}$ also vanishes for a pure Rashba system, as $\chi_{11}=3\chi_{21}$ and $\chi_{41}=0$. The dependence of $\chi_{\perp}^{\text{II}}$ on $\theta$ for different coupling strengths is illustrated in Figs. \ref{fig2}(b) and \ref{fig2}(e), corresponding to $\epsilon_F>0$ and $\epsilon_F<0$, respectively. For $\epsilon_F>0$, using the same parameters as those employed for $\chi_{\perp}^{\text{I}}$, we note that with increasing $\beta$, the magnitude of $\chi_{\perp}^{\text{II}}$ increases and exhibits a more pronounced anisotropic growth. We highlight two distinct behaviors of $\chi_{\perp}^{\text{II}}$: (i) there are no additional zeroes observed for any values of $\beta$, and (ii) in contrast to the case of $\chi_{\perp}^{\text{I}}$, the magnitude of $\chi_{\perp}^{\text{II}}$ does not show a drastic increase as $\beta$ approaches $\alpha$. This occurs because the BCP increases more rapidly as $\beta$ approaches $\alpha$, compared to the band velocity, which straightforwardly affects their respective contributions to the conductivity. 
  %When $\beta$ equals $\alpha$, $\chi_{\perp}^{\text{II}}$ becomes zero due to the equal and opposite contributions from both bands.
   The magnitude of $\chi_{\perp}^{\text{II}}$ is greater for $\epsilon_F<0$ when compared to the case of $\epsilon_F>0$. The magnitude and sign of $\chi_{\perp}^{\text{II}}$ also remain unchanged upon the interchange of $\alpha$ and $\beta$, along with a similar unitary transformation.

\begin{figure}[htbp]
\includegraphics[trim={0cm 0cm 0.0cm  0cm},clip,width=8.5
cm]{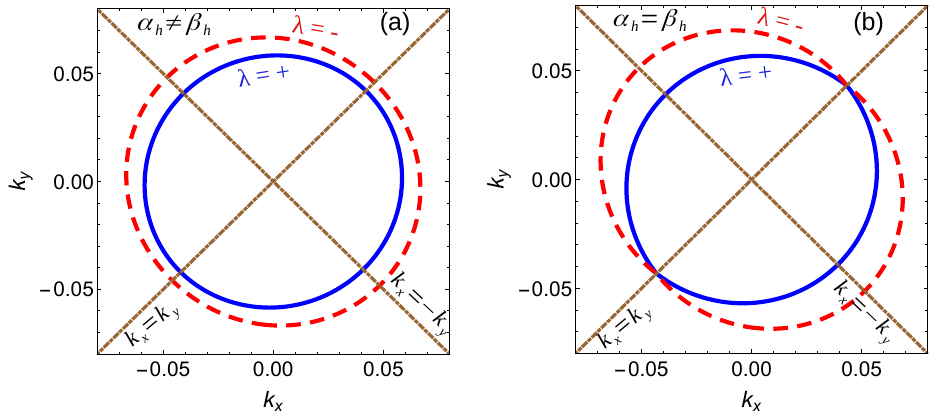}
\caption{Fermi contours along with the two mirror symmetric lines $k_y = \pm k_x$ of the 
2DHG with $k$-cubic Rashba-Dresselhaus spin-orbit couplings for
(a) $\alpha_h \neq \beta_h$ and (b) $\alpha_h = \beta_h$. Here, $k_x$ and $k_y$ are plotted in units of $k_h$. } 
\label{fig-contour-hole}
\end{figure}

  \begin {figure*}
    \centering
    \includegraphics[width=1.0\textwidth]{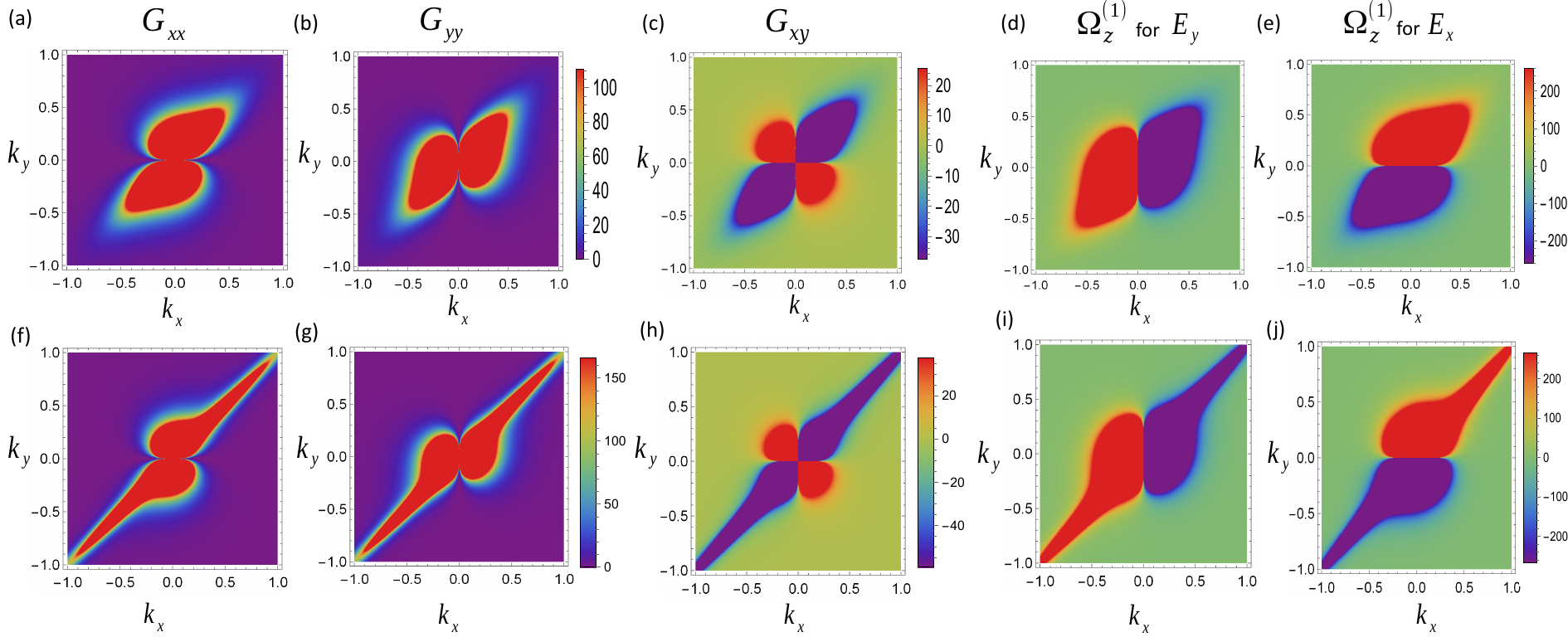}
 \caption{Distribution of the BCP tensors and the field-induced Berry curvature for + branch of a 2DHG with $k$-cubic  Rashba-Dresselhaus spin-orbit interactions: (top panel) ${\alpha_h}= 0.1$ eV nm$^3$ and ${\beta_h}=0.6 {\alpha_h}$, and  (bottom panel) ${\alpha_h}= 0.1$ eV nm$^3$ and ${\beta_h}=0.9 {\alpha_h}$. Here, (a)-(c) and (f)-(h) represent the density plots of the BCP tensor components (in units of $-e/\alpha_h {k_h^5}$), (d)-(e) and (i)-(j) represent the field-induced Berry curvature $\Omega_z^{(1)}$ (in units of $ {-e E/\alpha_h k_h^6}$) for two orientations of the electric field along the $y$ and  $x$ directions, respectively. In both panels, $k_x$ and $k_y$ are plotted in units of $k_h$. We consider $m_h= 0.41 m_0$, where $m_0$ is the free electron mass.} 
\label{fig3}
\end{figure*}
%  (in units of $e/{\alpha_h} {k_h^5}$)

   \textit{\textbf{Net transverse conductivity}} ($\chi_{\perp}$): We also explore the third-order transverse conductivity, which comprises two components proportional to $\tau$ and $\tau^3$, denoted as $\chi_{\perp}=\chi_{\perp}^{\text{I}}+\chi_{\perp}^{\text{II}}$. Extracting these two conductivities individually in an experimental setting proves challenging. Therefore, providing their combined contributions becomes a valuable approach at very low temperatures. However, the separation of these contributions has been demonstrated through temperature scaling analysis\cite{cwang}. We present the variation of $\chi_{\perp}/\chi_{1}$ as a function of $\theta$ for both $\epsilon_F>0$ and $\epsilon_F<0$ in Figs. \ref{fig2}(c) and (f). We have $\chi_{\perp}/\chi_{1}=\chi_{\perp}^{\text{I}}/\chi_{1}+(\chi_{\perp}^{\text{II}}/\chi_{2})(\chi_2/\chi_1)$ with $\chi_2/\chi_1=3.29$ for $\tau=1$ ps. When $\beta$ is significantly smaller than $\alpha$, the magnitude of $\chi_{\perp}^{\text{II}}$ surpasses that of $\chi_{\perp}^{\text{I}}$, resulting in the behavior of $\chi_{\perp}$ resembling that of $\chi_{\perp}^{\text{II}}$. When $\beta$ approaches values close to $\alpha$, both $\chi_{\perp}^{\text{I}}$ and $\chi_{\perp}^{\text{II}}$ become comparable. Consequently, we also observe additional zeros in the behavior of $\chi_{\perp}$, mirroring the pattern seen in $\chi_{\perp}^{\text{I}}$ for $\beta=5$.
   
   Based on our calculations, we  provide an estimate of the third-order Hall current that can potentially manifest during experimental observations. The third-order Hall current can be defined as $I= j^{(3)}_{\perp}l_0$, where $j^{(3)}_{\perp}=\chi_{\perp}E^3$ and $l_0$ represents the length of the sample. For an uniform electric field of $100$ V/cm, $l_0=1$ mm, $\tau=1$ ps, $\theta=\pi/2$, and utilizing system parameters such as $\alpha= 6\times 10^{-9}$ eV cm, $\beta= 1\times 10^{-9}$ eV cm and $\epsilon_{F}=4.27$ meV, the third-order  Hall current can be calculated as $I\sim15$ ${\mu}$A.
   
\section{Two-dimensional hole gas with $k$-cubic Rashba-Dresselhaus spin-orbit coupling}\label{section4}
The effective Hamiltonian of a heavy-hole gas with $k$-cubic RSOI and DSOI 
formed at the p-type III-V semiconductor heterostructures is given by\cite{loss,john1,mireles}
\begin{equation}
\centering
\begin{aligned}
H&=\frac{\hbar^2 k^2}{2m_h}+{i\alpha_h}\Big(k_-^3\sigma_+ -k_+^3\sigma_- \Big)\\
&-{\beta_h}\Big(k_- k_+ k_- \sigma_+ +k_+ k_- k_+ \sigma_- \Big),
\end{aligned}
\end{equation}
where $k_{\pm}=k_x \pm ik_y$, $\sigma_{\pm}=(\sigma_x \pm i\sigma_y)/2$, with $\sigma_i$'s as the Pauli spin matrices and $m_h$ is the effective heavy-hole mass. 
Also, $\alpha_h$ and $\beta_h$ are the strength of RSOI and DSOI, respectively. 
%The value of $\alpha_h$ typically falls within the range of 0.1 eV nm$^3$ for narrow-gap semiconductors, while $\beta_h$ is consistently smaller than $\alpha_h$. 
The energy spectrum is given by
\begin{equation}
\epsilon_{\lambda}\mathbf{(k)}=\frac{\hbar^2 k^2}{2m_h}+\lambda k^2\sqrt{{(\alpha_h k_x-\beta_h k_y)}^2+{(\alpha_hk_y-\beta_h k_x)}^2},
\end{equation}
\begin{figure}[htbp]
\includegraphics[trim={0cm 5.2cm 0.0cm  0cm},clip,width=8.3
cm]{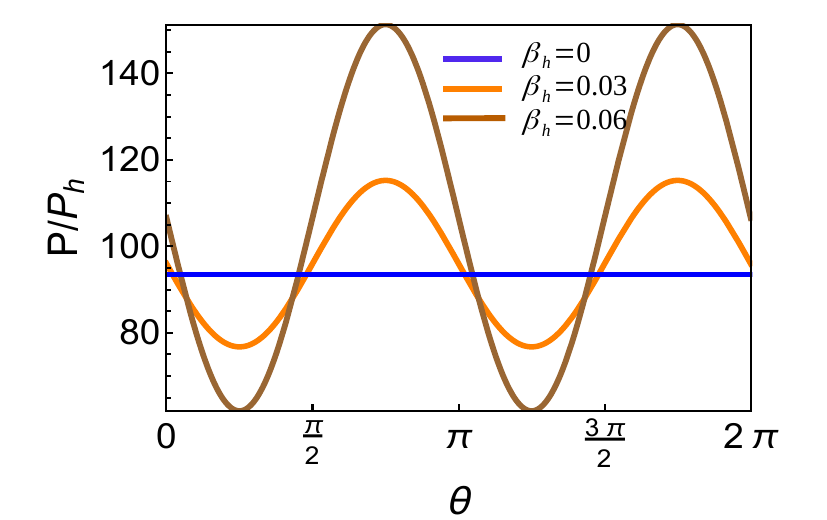}
\caption{Polarization (in units of $P_h=e^2 E/\epsilon_h$) for a hole gas  with angle $\theta$ for different Dresselhaus coupling strengths (given in units of  eV nm$^3$) at a fixed Rashba coupling strength $\alpha_h=0.1$ eV nm$^3$. The other parameters used are charge carrier density $n_h=2\times 10^{15}$ m$^{-2}$ and ${m_h}= 0.41 {m_0}$\cite{alestin}. }  
\label{pol-hole}
\end{figure}
where $\lambda=\pm$ denotes the two dispersive branches. The corresponding eigenspinors can be calculated as $\vert{u}_{\lambda}\rangle=(1/\sqrt{2})[1 \hspace{0.25 cm}\lambda  \e^{i(2\phi-\phi')}]^{T} $, where  $\phi'=\tan^{-1}[({\alpha_h k_x-\beta_h k_y})/({\alpha_h k_y-\beta_h k_x})]$ with $k_x=k \cos\phi$ and $k_y=k \sin\phi$. The spin splitting energy between the two branches, $\epsilon_g(\mathbf{k})=\epsilon_{+}(\mathbf{k})-\epsilon_{-}(\mathbf{k})= 2k^2\Delta_{\mathbf{k}}$, with  $\Delta_{\mathbf{k}} \equiv\Delta=\sqrt{(\alpha_h k_x-\beta_h k_y)^2+(\alpha_hk_y-\beta_h k_x)^2}$. In polar form, it can be expressed as $\epsilon_g(\mathbf{k})=2k^3\vartheta(\phi)$, where $\vartheta(\phi)\equiv \vartheta =\sqrt{\alpha_h^2+\beta_h^2-2\alpha_h \beta_h \sin2\phi}$.
It is to be noted that the lower branch of the Hamiltonian is valid for the wave numbers $k\leq\hbar^2/(2 m_h \vartheta)$.
The maximum value of $\epsilon_g(\mathbf{k}) $ at $\phi=3\pi/4$ and $7\pi/4$  is 
$ 2k^3(\alpha_h+\beta_h)$, and the minimum value of $\epsilon_g(\mathbf{k}) $ at $\phi=\pi/4$ and $5\pi/4$ is $2k^3|\alpha_h-\beta_h|$. 
These values of $\phi$ also coincide with the symmetry lines $k_x \pm k_y=0$ of the system. 
There is a line degeneracy along the symmetry line $k_y - k_x = 0$ for $\alpha_h = \beta_h$ case
as shown in Fig. \ref{fig-contour-hole}.

\begin {figure*}
    \centering
    \includegraphics[width=1\textwidth]{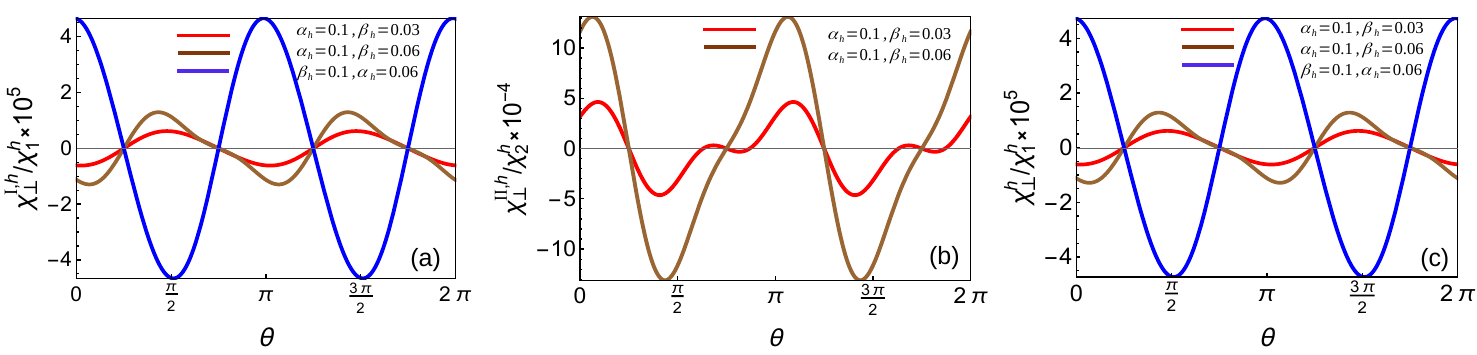}
 \caption{(a)-(c) Variation of the transverse third-order conductivities for the heavy-hole gas with $k$-cubic Rashba-Dresselhaus spin-orbit interactions as a function of the angle $\theta$ between the electric field and the $x$-axis. The conductivities $\chi_{\perp}^{\text{I},h}$ and $\chi_{\perp}^{\text{II},h}$ represent the transverse third-order conductivities of the hole gas proportional to $\tau$ and $\tau^3$, respectively. The total transverse conductivity is given by $\chi_{\perp}^{h} = \chi_{\perp}^{\text{I},h} + \chi_{\perp}^{\text{II},h}$. In (b), $\chi_{\perp}^{\text{II},h}$ ($\alpha_h=0.1$ and $\beta_h=0.06$) = $\chi_{\perp}^{\text{II},h}$ ($\beta_h=0.1$ and $\alpha_h=0.06$). The normalization parameters for conductivities, $\chi_{\perp}^{\text{I},h}$ and $\chi_{\perp}^{\text{II},h}$, are given by $\chi_{1}^{h} = {\tau e^4 m_h^5 \alpha^4 }/{\hbar^{12} }$ and $\chi_{2}^{h}={e^4 \tau^3}/{m_h \hbar^2}$, respectively. The value of  $\alpha_{h}$ and $\beta_h$ are given in units of  eV nm$^3$. The parameters used are the same as in Fig. \ref{pol-hole}. }  
\label{fig4} 
\end{figure*}

   The analytical derivation of wave vectors is not feasible for the anisotropic hole system. Hence, we numerically evaluate the wave vectors by solving the cubic equation, $\hbar^2 k^2/2m_h + \lambda k^3 \vartheta - \epsilon = 0$. However, when $\beta_h$ is set to zero, exact expressions for the Fermi wave vectors can be obtained analytically\cite{john1}.  The scaled wave vector and energy are defined as $\tilde{k}_h=k/k_h$ and $\tilde{\epsilon}=\epsilon/\epsilon_h$, where $k_h=\hbar^2/({m_h \alpha_h})$ and $\epsilon_h=\alpha_h {k_h^3}$.
   
The Berry connection for the system can be calculated 
as $\mathbf{A}_{\mathbf{k}}=\frac{\delta}{2 k^2 {\Delta}^2}(k_y \hat{x}-k_x \hat{y})$, where $\delta=[(3\alpha_h^2+\beta_h^2)(k_x^2+k_y^2)-8\alpha_h\beta_h k_x k_y]$. 
%The corresponding Berry phase is given by $\Gamma_{\lambda}= 2\pi+\pi{\text{sgn}(\alpha^2-%\beta^2)}$. 
%Unlike the electron gas case with both SOIs, the Berry connection and the Berry phase for the hole gas do not vanish when $\alpha_h=\beta_h$. 
The Berry curvature is zero, which leads to the absence of linear and second-order Hall responses, making the third-order Hall response dominant for the hole system as well. To calculate the third-order conductivity, one can evaluate the different components of the BCP tensor for the system as
\begin{equation} \label{Gabh}
G_{ab}^{\lambda}=-\lambda\frac{ e\delta^2 }{4\Delta^5{k}^6}\left (\begin{matrix}
   {k_{y}^{2}}  & {-k_{x} k_{y}} \\
  
  {-k_{x} k_{y}} & {k_{x}^{2}}
\end{matrix} \right).
\end{equation}
Similar to the electron case, Eq. (\ref{Gabh}) is not valid for $\alpha_h =\beta_h$ because
of the presence of the line degeneracy along symmetry line $k_y-k_x =0$.
The distribution of the BCP tensor components in the $k_x$-$k_y$ plane for $\alpha_h=0.1$ eV nm$^3$ and $\beta_h=0.6\alpha_h$ is plotted in Figs. \ref{fig3}(a)-\ref{fig3}(c). The diagonal components $G_{xx}$ and $G_{yy}$ show a dumbbell-like structure, whereas $G_{xy}$ exhibits quadrupole-like features. On applying an in-plane electric field, the second-order energy correction and the field-induced Berry curvature can be obtained as
%\begin{equation}
%\epsilon_{\lambda}^{(2)}=\lambda\frac{e^2{\delta^2}}{8 k^6 \Delta^5}(\mathbf{E}\times\mathbf{k})^{2},
%\end{equation}
%and
%\begin{equation}
%\boldsymbol{\Omega}^{(1)}_{\lambda}(\mathbf{k})=\lambda\frac{e{\delta^2}}{ k^6 \Delta^5}(\mathbf{E}\times\mathbf{k}).
%\end{equation}
\begin{equation}
\epsilon_{\lambda}^{(2)}=\lambda\frac{e^2{\delta^2}}{8 k^6 \Delta^5}(\mathbf{E}\times\mathbf{k})^{2}\hspace{0.3cm} \text{and}  \hspace{0.3cm}
\boldsymbol{\Omega}^{(1)}_{\lambda}(\mathbf{k})=-\lambda\frac{e{\delta^2}}{ k^6 \Delta^5}(\mathbf{E}\times\mathbf{k}).
\end{equation} 
%\begin{equation}
%\begin{aligned}
%\hbar v_{x,{\lambda}}&=\frac{\hbar^2 k_x}{m_h^*}+\lambda\Bigg[\frac{3k_x(\alpha_h^{2}+\beta_h^2)k^2 -2\alpha_h\beta_h k_y(5k_x^2+k_y^2)}{\Delta}\Bigg],\\
%\hbar v_{y,{\lambda}}&=\frac{\hbar^2 k_y}{m_h^*}+\lambda\Bigg[\frac{3k_y(\alpha_h^{2}+\beta_h^2)k^2 -2\alpha_h\beta_h k_x(5k_y^2+k_x^2)}{\Delta}\Bigg],
%\end{aligned}
%\end{equation}
Similar to the electron gas case, we observe that $\Omega_{z}^{(1)}$ exhibits a dipole-like structure with its orientation changing relative to the electric field direction, as depicted in Figs. \ref{fig3}(d)-\ref{fig3}(e). When $\beta_h$ is zero, the lobes align precisely along the $x$ and $y$ axes. As we increase $\beta_h$, anisotropy is introduced into the system, causing the lobes in the BCP components and $\Omega_{z}^{(1)}$ to align within the $x$-$y$ plane. Further increase of $\beta_h$ results in the stretching of lobes, as shown in Figs. \ref{fig3}(f)-\ref{fig3}(j).

\subsection{ Polarization}
Similar to the electron case, we obtain an analytical expression 
for the electric polarization of 2DHG with $k$-cubic RSOI $(\beta_h=0)$,
\begin{equation}
P=\frac{3{P_h}}{2\pi}\Big[\frac{3(1+\sqrt{1-16\pi {n}_h l_h^2}) - 
32 \pi n_h l_h^2}{16 \pi n_h l_h^2 (1-16 \pi n_h l_h^2 )^{3/2}}\Big],
\end{equation}
where $P_h=e^2 E/\epsilon_h$ and $ l_h = m_h\alpha_h/\hbar^2$. 
%$\tilde{n}_h=( \pi{m_h^2} {\alpha_h^2}/\hbar^4)n_h$. 
For $\alpha_h = 0 $ and $\beta_h \neq 0 $, the polarization is reduced by a factor of nine.
Here as well, polarization remains constant with $\theta$ when either one of the spin-orbit couplings is absent, for similar reasons as specified in the electron case.
The variation of polarization with $\theta$ in the presence of both the couplings is depicted in Fig. \ref{pol-hole}. The polarization increases with 
$\beta_h$, while decreases with the Fermi energy. 
When both $\alpha_h$ and $\beta_h$ are nonzero, the integration of $G_{xx}$ and $G_{xy}$ yield positive and negative values, respectively. Thus, the maximum of polarization is observed at $\theta=3\pi/4$ and $7\pi/4$ and minimum at $\theta=\pi/4$ and $5\pi/4$. This is in contrast to the electron case. 
    
  For a positive Fermi energy, the polarization of a $k$-linear electron gas with RSOI and DSOI is of an order of magnitude smaller than that for a hole gas with $k$-cubic couplings.

\subsection{ Third-order transverse conductivity}

The $k$-cubic Rashba-Dresselhaus system acquires the same form of $\chi_{\perp}$ as described in  Eq. (\ref{chi-perp}), owing to the same symmetry lines $k_x\pm k_y=0$. Next, we discuss the contribution of $\chi_{\perp}$ proportional to $\tau$ and $\tau^3$ given by Eqs. (\ref{tau}) and (\ref{tau3}) for the hole system.

\textit{\textbf{$\tau$-scaling conductivity}} ($\chi_{\perp}^{\text{I},h}$): We evaluate $\chi_{\perp}^{\text{I},h}$ numerically for different values of $\alpha_h$ and $\beta_h$, and its variation with respect to $\theta$ is depicted in Fig. \ref{fig4}(a). In our calculations, we consider the parameters representing p-type InAs heterostructures\cite{alestin}: hole density $n_h=2\times10^{15}$ m$^{-2}$ and $m_h = 0.41 m_0$, and $\alpha_{h}=0.1$ eV nm$^3$, while varying $\beta_h$. In an isotropic cubic Rashba system, $\chi_{\perp}^{\text{I},h}$ is zero since $3\chi_{21}=\chi_{11}$ and $\chi_{41}=0$. However, when a finite small  value of $\beta_h$ is introduced, $\chi_{\perp}^{\text{I},h}$ becomes finite and exhibits a significant dependence on the $\cos2\theta$ term. It is important to note that as we increase $\beta_h$ from 0.1$\alpha_h$ to 0.5$\alpha_h$, the curve of $\chi_{\perp}^{\text{I},h}$ follows qualitatively a similar pattern but with an increased magnitude. This happens because the BCP is proportional to $\delta^2$ and more specifically, the coefficient associated with $\alpha_h$ is three times that of $\beta_h$. Therefore, as $\beta_h$ is increased, the impact on $\delta^2$ is less pronounced compared to changes in $\alpha_h$, resulting in the observed pattern of $\chi_{\perp}^{\text{I},h}$ with a higher magnitude but similar overall shape. As $\beta_h$ is further increased, anisotropic curves emerge from the interplay between the coefficients of $\sin4\theta$ and $\cos2\theta$. Similar to the electron scenario, we notice additional angles at which $\chi_{\perp}^{\text{I},h}$ vanishes, beyond those dictated by the system's inherent symmetry. Note that these angles of additional zeros depend on the system parameters. The positions of maxima and minima shift as one varies $\alpha_h$ and $\beta_h$, emphasizing their
dependence on system parameters.
     
%     In the special case where $\alpha_h=\beta_h$, the energy bands $\epsilon_{\lambda}(\mathbf{k})$ become degenerate at $\phi=\pi/4$ and $5\pi/4$. The second-order semiclassical formalism, which is used to calculate the third-order conductivity induced by BCP (where $\epsilon_{\lambda}^{(2)} \propto G_{ab}$), is primarily applicable to the nondegenerate bands. Consequently, the calculation of $\chi_{\perp}^{\text{I},h}$ at the degenerate points cannot be carried out using this formalism. Thus the results for $\alpha_h = \beta_h$ are not presented here.
     
Upon applying a unitary transformation $U$ similar to that used for the electron case and interchanging the values of $\alpha_h$ and $\beta_h$, the transformed Hamiltonian no longer remains invariant. The perturbed velocity resulting from $\epsilon_{\lambda}^{(2)}$ changes under such transformations. Therefore, the third-order conductivity ($\propto \tau$) ceases to remain invariant under $\alpha_h \leftrightarrow \beta_h$, as evident in Fig. \ref{fig4}(a).
     
\textit{\textbf{$\tau^3$-scaling conductivity}} ($\chi_{\perp}^{\text{II},h}$): The variation of $\chi_{\perp}^{\text{II},h}$ as a function of $\theta$ for the same set of parameters is shown in Fig. \ref{fig4}(b). We find that the $\chi_{\perp}^{\text{II},h}$ vanishes for an isotropic Rashba system ($\beta_h=0$), for the same underlying reason observed for $\chi_{\perp}^{\text{I},h}$. The magnitude of $\chi_{\perp}^{\text{II},h}$ increases with the $\beta_h$, while keeping $\alpha_h$  fixed. When $\alpha_h=\beta_h$, $\chi_{\perp}^{\text{II},h}$ becomes zero due to equal and opposite contributions from both the branches. The magnitude and sign of $\chi_{\perp}^{\text{II},h}$ remains unchanged upon interchanging $\alpha_h$ and $\beta_h$ is a direct consequence of its origin in the unperturbed velocity, which remains insensitive to such transformations.
  
  \textit{\textbf{Net transverse conductivity}} ($\chi_{\perp}^{h}$): In Fig. \ref{fig4}(c), we present the variation of the net contribution $\chi_{\perp}^{h}$ arising from $\tau$ and $\tau^3$. It is worth noting that the magnitude of $\chi_{\perp}^{\text{II},h}$ is smaller than that of $\chi_{\perp}^{\text{I},h}$ for a hole gas. As a result, the behavior of $\chi_{\perp}^{h}$ exhibits similarity to that of $\chi_{\perp}^{\text{I},h}$. Like $\chi_{\perp}^{\text{I},h}$ and $\chi_{\perp}^{\text{II},h}$, $\chi_{\perp}^{h}$ varies with $\theta$ with a period of $\pi$.
   
    For the Hall setup with the same parameters as those employed for the electron case and the system parameters specified as $\alpha_h= 0.1$ eV nm$^3$ and $\beta_h= 0.3 \alpha_h$, the estimated third-order  Hall current for the hole gas with $k$-cubic RSOI and DSOI is $I_h\sim12$ ${\mu}$A.

\section{conclusion}\label{section5}
In this study, we investigated the electric polarization and third-order Hall response in a 2D electron/hole gas with $k$-linear/$k$-cubic RSOI and DSOI present at III-V semiconductor heterostructures. 
%The Berry curvature of such time-reversal symmetric system is zero. 
We have obtained the analytical expressions of the BCP tensors and the field-induced Berry curvature. We have also obtained  analytical expressions for the BCP-induced electric polarization when
either Rashba or Dresselhaus spin-orbit interaction is present. 
The electric polarization decreases with an increase in the Fermi energy, while it increases with the Dresselhaus coupling for a given Rashba coupling. We find that the polarization is sensitive to the orientation of the electric field when both Rashba and Dresselhaus spin-orbit couplings are present. For the Fermi energy above the BTP, the polarization of 2DEG with Rashba-Dresselhaus spin-orbit interaction is of an order of magnitude smaller than that for the 2DHG.

The Berry curvature of such time-reversal symmetric system is zero. 
Consequently, both the linear Hall effect and the second-order nonlinear Hall effect (induced by the Berry curvature dipole) are absent. 
As a result, the third-order response becomes the dominant Hall effect in these systems. 
Using second-order semiclassical formalism, we have computed the third-order conductivity induced by the BCP, which is linearly proportional to $\tau$. Furthermore, we extended our analysis to the third-order conductivity stemming from band velocity, which is cubic in $\tau$, and also studied their cumulative effects.

Next, we examine the effect of an in-plane electric field and calculate the transverse third-order conductivities, namely $\chi_{\perp}^{\text{I}}$, $\chi_{\perp}^{\text{II}}$, and $\chi_{\perp}$ ($\chi_{\perp}^{\text{I},h}$, $\chi_{\perp}^{\text{II},h}$, and $\chi_{\perp}^h$) for electron (hole) system, while varying the coupling strengths. We find that these conductivities vanish along or perpendicular to the symmetry lines $k_x \pm k_y = 0$ of the system, specifically at odd multiples of $\pi/4$. These responses exhibit $\pi$ periodicity with respect to the direction of the electric field. In the absence of either coupling, energy dispersions become isotropic with concentric circular Fermi contours. As a result, all  contributions involving $\tau$ and $\tau^3$ to transverse third-order conductivities vanish across all angles. Thus it is the interplay between RSOI and DSOI that engenders to finite transverse third-order conductivity.
  
   For the case of an electron gas with $k$-linear RSOI and DSOI, we find that $\chi_{\perp}^{\text{I}}$ exhibits a smaller magnitude compared to $\chi_{\perp}^{\text{II}}$ for $\beta<\alpha$. However, the magnitude of $\chi_{\perp}^{\text{I}}$ significantly increases as $\beta$ approaches proximity to $\alpha$ in comparison to $\chi_{\perp}^{\text{II}}$. This is attributed to the nature of BCP and the band velocity. 
%The $\chi_{\perp}^{\text{II}}$ disappears at $\alpha=\beta$ due to counterbalancing contributions %from both the bands. 
The magnitudes of conductivities are larger for $\epsilon_F < 0$ than for $\epsilon_F > 0$. 
%Unlike the linear spin Hall current, 
The third-order conductivity ($\chi_{\perp}^{\text{I}}$ and $\chi_{\perp}^{\text{II}}$) remains invariant under the interchange of $\alpha$ and $\beta$. This is due to the invariance of both the unperturbed velocity and the velocity resulting from the second-order energy correction when $\alpha$ and $\beta$ are exchanged. 
   
  Comparing a 2DHG with $k$-cubic RSOI and DSOI to the $k$-linear electron model, we observe that the magnitude of $\chi_{\perp}^{\text{I},h}$ is larger  compared to $\chi_{\perp}^{\text{II},h}$. Therefore, $\chi_{\perp}^{h}$ shows a curve similar to that of $\chi_{\perp}^{\text{I},h}$.
  % At $\alpha_h=\beta_h$, $\chi_{\perp}^{\text{II},h}$ becomes zero because both branches contribute equally and oppositely. 
  When $\alpha_h$ and $\beta_h$ are exchanged, $\chi_{\perp}^{\text{I},h}$ undergoes a change due to the sensitivity of the BCP tensor to such transformations. In contrast, $\chi_{\perp}^{\text{II},h}$ remains invariant since the unperturbed velocity remains constant.

\begin{center}
	{\bf ACKNOWLEDGEMENT}
\end{center}
We would like to thank Bashab Dey for
useful discussions.

\end{document}